\documentclass[twocolumn,showpacs,preprintnumbers,amsmath,amssymb]{revtex4}
\usepackage{graphicx}% Include figure files
\usepackage{dcolumn}% Align table columns on decimal point
\usepackage{bm}% bold math

\begin{document}

\preprint{Phys Rev E}

\title{From regular lattice to scale free network\\
- yet another algorithm}% Force line breaks with \\

\author{Danuta Makowiec}
 \email{fizdm@univ.gda.pl}
\affiliation{%
Institute of Theoretical Physics and Astrophysics\\
Gda\'nsk University, Poland }%

\date{\today}% It is always \today, today,

\begin{abstract}
The Watts-Strogatz algorithm transferring a regular lattice to the small world network is modified by introducing preferential rewiring constrained by connectivity demand. The probability to link to/ unlink form a node is dependent on a vertex degree and adjusted by some threshold. For each threshold value there exists a probability at which the resulting stationary network has degree distribution with power-law decay in large interval of degrees.
\end{abstract}

\maketitle

\section{\label{sec:Intro} Introduction}
Networks are used extensively to study and describe topology in various complex real systems.
From communication networks like World Wide Web, through nets of social relations like network of friends or collaborating scientists, to biological systems like  protein network, neural network or cell metabolism --- a similar structural organization has been discovered,  see \cite{AlbertBarabasi02,DorogovtsevMendes02,Newman03} for review of data analysis and bibliography. 

The simplest and the most intensely studied characteristic of a network is degree of a vertex. Degree, $k$, of a vertex is the total number of its connections. A physicist would say degree is actually the number of nearest neighbors of a vertex. Distribution of vertex degrees of an entire network is  the network's basic statistical characteristic. Although the degree of a vertex is a local quantity, the degree distribution often determines some  important global  features. For example, in case of a regular lattice (a network explored in many physical applications) the degree distribution is concentrated at a single value what exactly proves the homogeneity of the network topology.

The studies of other topologies began with the random graph theory of Erdos and Renyi \cite{ErdosRenyi}.  The proposition of Watts and Strogatz \cite{WattsStrogatz98} that followed, called small world network, captures  best the features of both regular lattice and random graph. The algorithm may be summarized as follows: begin with a regular lattice, then rewire an edge that is chosen with some probability $p$. The crucial properties of the resulting network are that it is highly clustered like a regular lattice, and at the same time it has  small path lengths like a random graph. 

The topology of some graphs found in technology such as electronic circuits, where  electronic components are represented as vertices  and the wires are edges, can be viewed as a class of such networks \cite{Cancho}. The magnetic materials, including Ising models, based on small world networks are  investigated widely in the aim of constructing magnetic nanomaterials \cite{Ising,Makowiec04}. The neural networks may also be considered as kind electronic circuits \cite{Fernandez,Mathias} according to the idea of `save wiring' as an organizing principle of the brain \cite{Cherniak}. 

The degree distribution of small world network is similar to a random graph. It has pronounced peak at the average degree $<k>$ and decays exponentially for large $k$. Thus the topology of the network is relatively homogeneous. However, the degree distribution of most real networks significantly deviates from such distribution. For large number of real networks  degree distribution has a power-law form:
\begin{equation}
 P(k) \propto k^{-\gamma}
 \label{gamma}
\end{equation}
where $\gamma$ is the degree exponent. A network with the property (\ref{gamma}) is called scale free network. In a scale free network, the smaller value of $\gamma$, the greater role of, so-called hubs: vertices with large degree. Especially, if $\gamma <3$ then the surprising network features such as robustness against accidental failure of a vertex occur \cite{AlbertJeongBarabasi}.

Barabasi and Albert proposed the method for constructing a network with the power-law degree distribution \cite {BA99}. The two basic features of the model are: the network growth (in each time step a new vertex is attached to a graph), and the preferences in the attachment (a vertex is in favor according to its degree). Dorogovtsev and Mandes \cite{DM00} proved that aging (a preference growth based on the age of the node) could lead to scale free network. Dangalchev \cite{Dangalchev} implied that the scale free topology can be restored in a model with a static number of nodes. However, the organizing principles that govern and are common in a wide range of complex networks are far from being fully understood \cite{MenezesBarabasi}. 

In the following paper we present the effect to the degree distribution of adding preferences to rewiring in Watts-Strogatz algorithm.  In analogy to the small world algorithm, we start from a regular lattice (a square lattice) and set the preferences in rewiring to be related to a vertex degree and  adjusted by some threshold. In addition, we set the constraint of preserving the connectivity of a graph. The resulting network at properly adjusted parameters has the  degree distribution  of power-law type in the wide interval. The degree exponent $\gamma$ is less or close  to $2$. Then in a tail, the distribution becomes an exponential one. 

The algorithm is presented in Section 2. Section 3 contains results from simulations. The planned development of the model is proposed in  Section 4.

\section{\label{sec:Alg}The algorithm}

The vertices of the lattice are numbered by: $1,2,\dots, N$. Each edge is characterized by the two numbers $from$ and $to$ representing two vertices linked by this edge. The graph is represented as the vector of size $N$ of lists of vertices  --- neighbors of subsequent vertices. The graph is not considered as directed, though it can easily be adapted to a directed one.

There are two basic  parameters in the model:\\
 --- probability to rewire an edge: $p$\\
 --- threshold to set in favor a vertex  in the preference function: $T$

\begin{figure}
\includegraphics[width=0.40\textwidth]{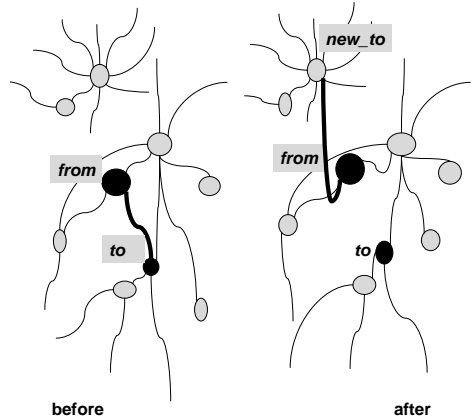}
\caption{\label{fig1} An edge linking vertices $from$ and  $to$ ( before graph) is rewired to the edge between vertices $from$ and $new\_to$ (after graph)}.
\end{figure}

At each time step, N vertices are considered for rewiring. A vertex for rewiring is selected  at random. 
The basic procedure, called, EdgeEvolution, requires three parameters: $from$, $p$ and $T$. The procedure involves considering changes in all edges linked to a vertex $from$.  The rewiring an edge means change a  present value $to$ to some other. The steps in the procedure are described below; the illustration of the `before and after' is shown in Fig.~\ref{fig1}.

{\it EdgeEvolution ( $from$, $p$, $T$) :}
\begin{enumerate}
\item[(a)]for each vertex $to$ from the list of edges of the vertex $from$ do
\item[(b)]\hspace{0.5cm} if degree $to$ is greater than 1
\item[(c)]\hspace{0.5cm}then
\item[(d)]\hspace{1.0cm}let $\xi_1\in [0,1]$ be a random number 
\item[(e)]\hspace{1.0cm}if 
                 $$\xi_1 <\frac{pT}{{\rm degree\  of\ }  to}$$
\item[(f)]\hspace{1.0cm} then accept vertex $to$ for $unlinking$
\item[(g)]for each vertex $to$ accepted for $unlinking$
\item[(h)]\hspace{0.5cm}let $new\_to\in \{ 1,2,\dots, N\}$ be a random   number  but $new\_to \neq from$
\item[(i)]\hspace{0.5cm}let $\xi_2\in [0,1]$ be a random number
\item[(j)]\hspace{0.5cm}if 
                 $$\xi_2 <\frac{{\rm degree\  of\ }  new\_to}{T}$$
\item[(k)]\hspace{0.5cm}then accept $new\_to$: 
$$edge(from,to):= edge(from, new\_to)$$
\item[(l)]\hspace{0.5cm}otherwise go to (h)

\end{enumerate}

Remarks:\\
--- The algorithm conserves both the number of vertices and the number of edges. Steps (a)--(f) prepare the list of edges of the vertex $from$ to be rewired, while steps (g)--(l) fix new connections. Each edge accepted for rewiring has to be rewired (see the loop(l)).\\
--- The preferences (e) and (j) are chosen to prevent unlinking and favor  linking to  vertices with degrees greater than the threshold value.\\
--- The condition (b) is crucial to conserve connectivity of a graph. Any vertex with the degree equal to 1 cannot be unlinked. Otherwise we face the problem of dramatically increasing number of isolated nodes.\\
--- A randomly selected  new vertex for linking to must be different from the number $from$ to avoid loops in a graph, line (h).

The update of  information regarding the vertex degree is made after each time step. Therefore, if the network is evolving with small $p$, then the evolution may be seen as asynchronous, while with large $p$, the evolution may look as if performing  synchronously.

\section{Results}

We test by simulations the algorithm on a square lattice with $L=100$ and $L=200$. It means that number of vertices considered is $N=10^4, 4*10^4$, respectively. The number of edges is $4N$. We   observe no difference in results obtained for different lattice sizes.

\begin{figure}
\includegraphics[width=0.4\textwidth]{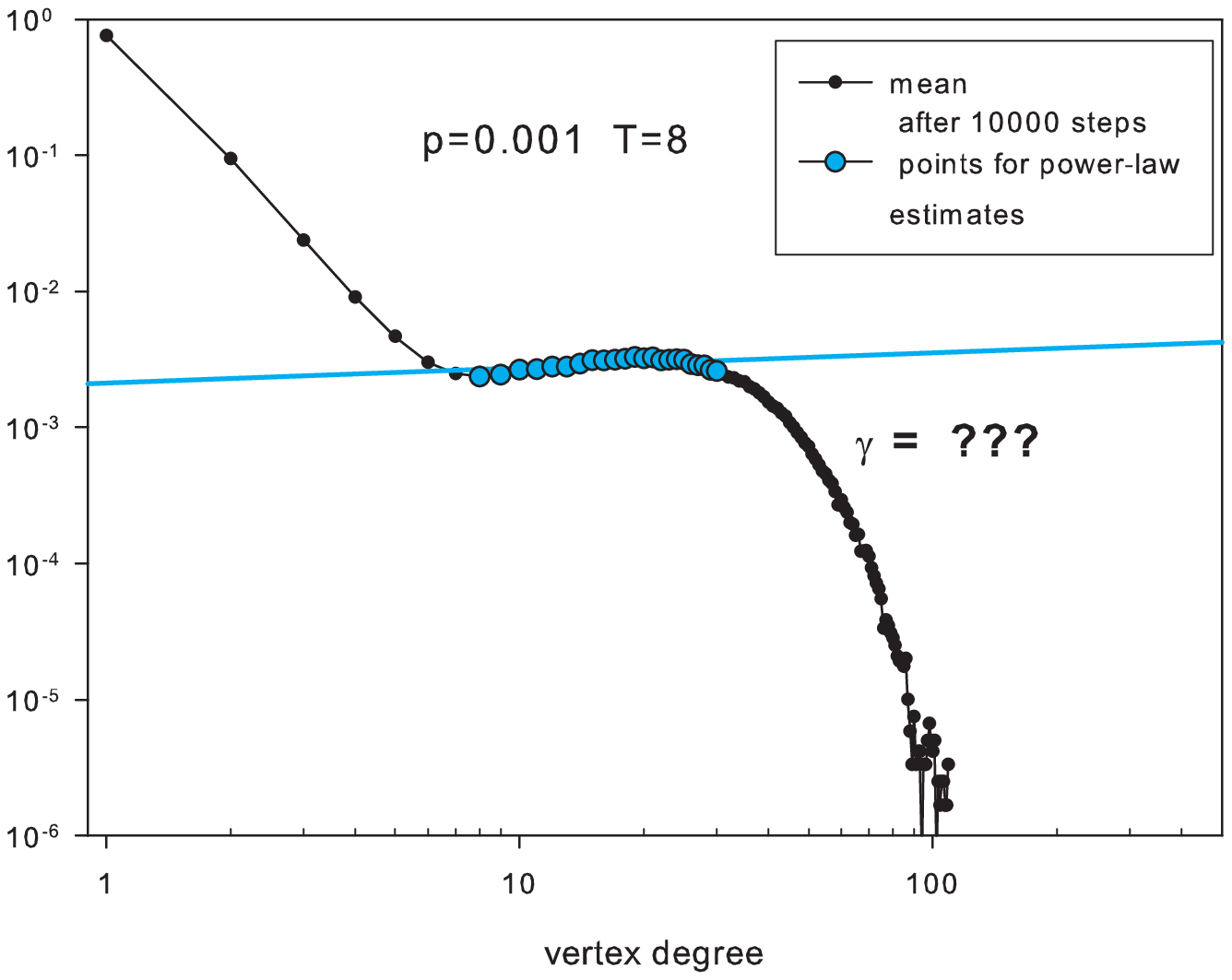}
\includegraphics[width=0.4\textwidth]{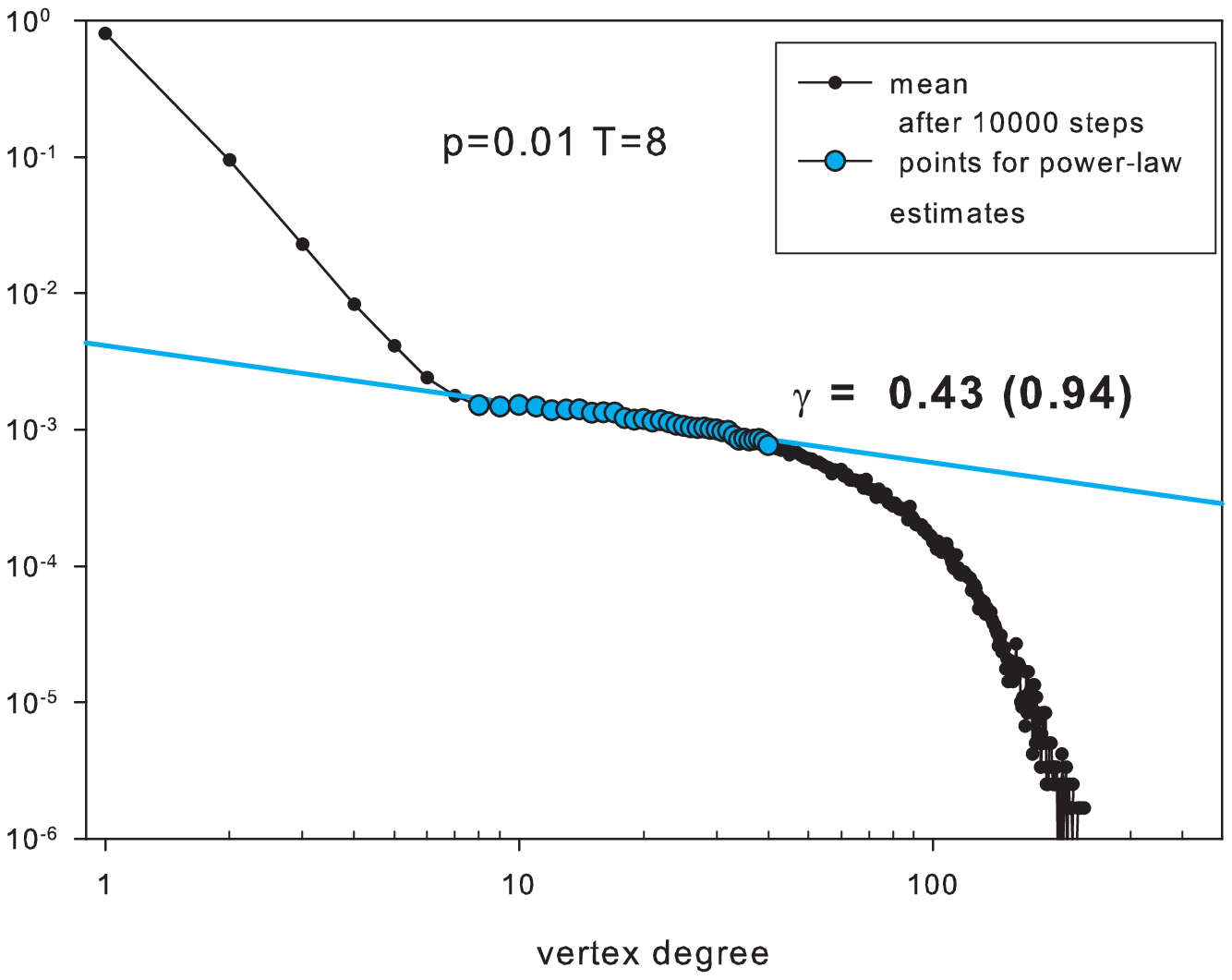}
\includegraphics[width=0.4\textwidth]{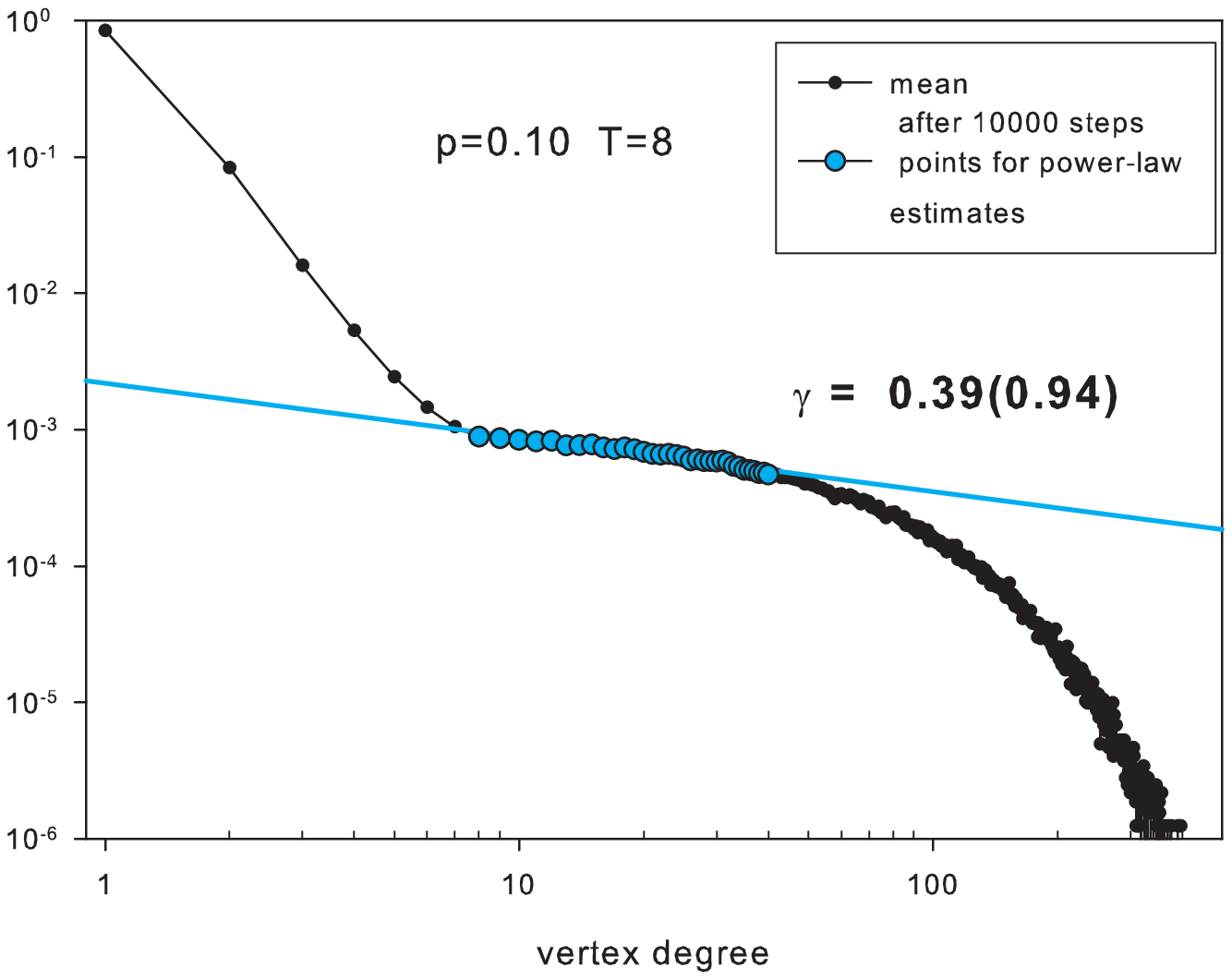}
\includegraphics[width=0.4\textwidth]{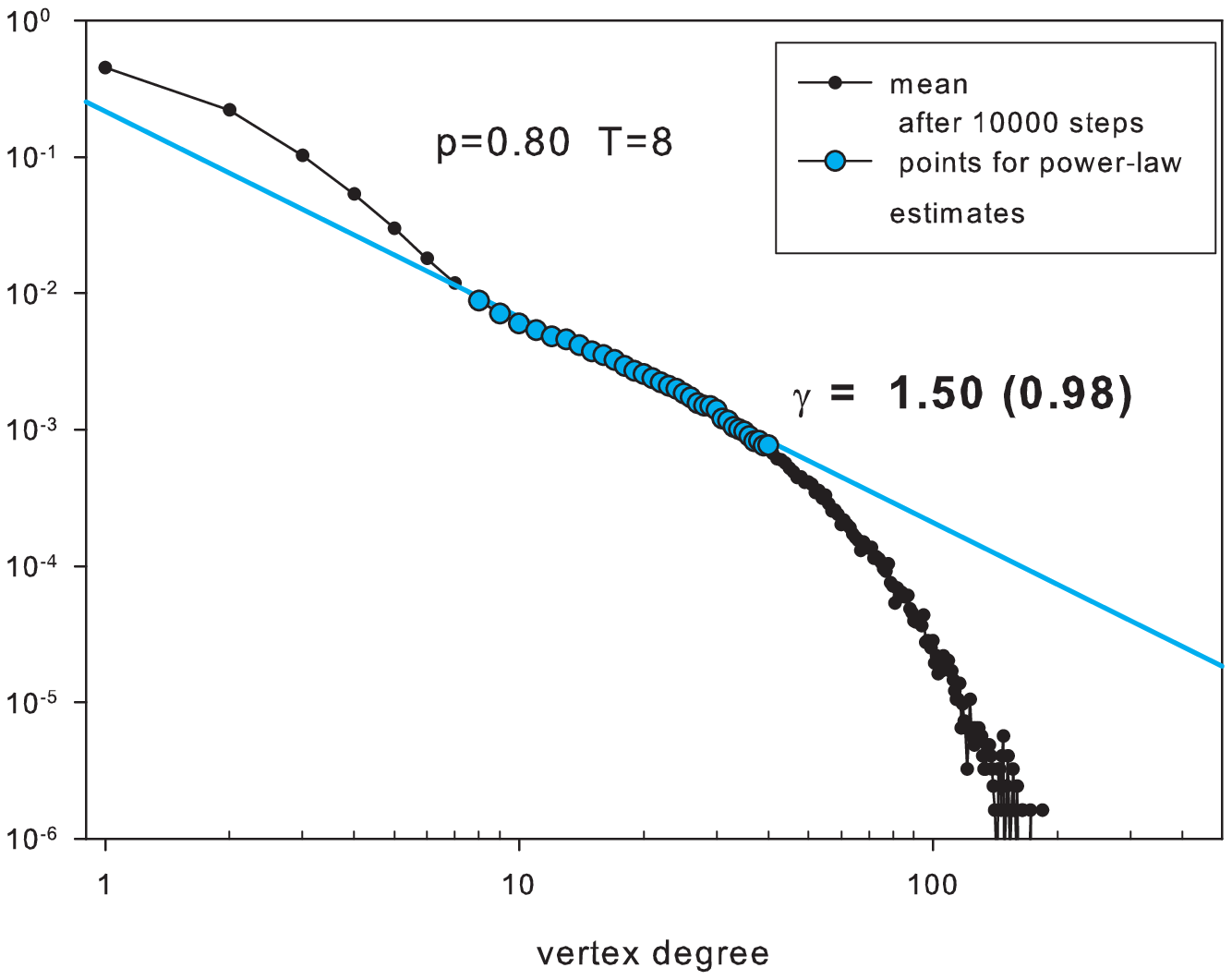}
\end{figure}
\begin{figure}
\includegraphics[width=0.4\textwidth]{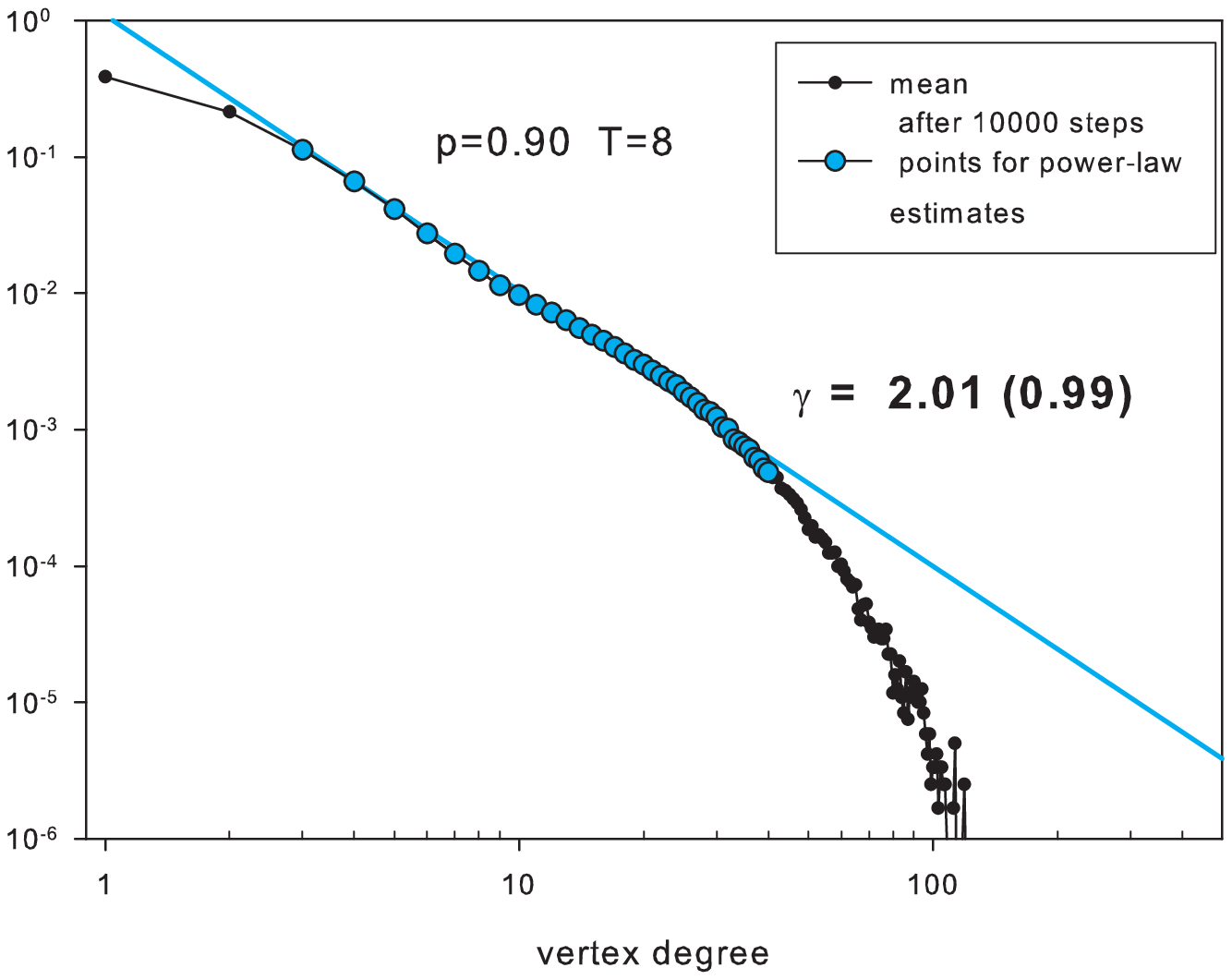}
\includegraphics[width=0.4\textwidth]{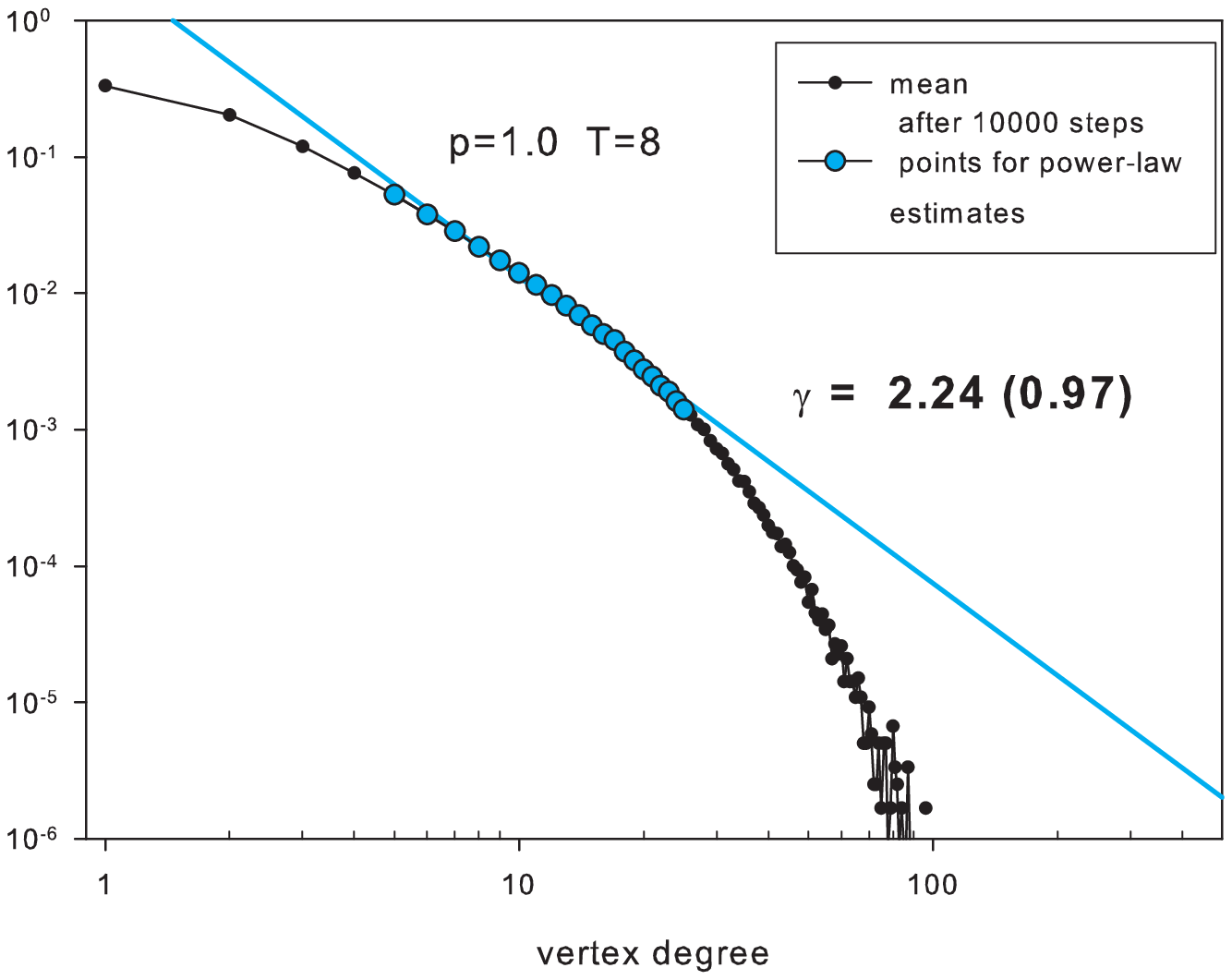}
\caption{\label{fig2} On subsequent figures the log-log plots of the degree distributions with $T=8$ and different $p$ are presented. The degree exponent values are calculated for points marked by gray big dots and are written at the plots together with their Pearson coefficients $r^2$.}
\end{figure}

For any $p$ and $T$ in a few time steps the initial $\delta$-like degree distribution centered at $k=4$ spreads to some peaked distribution  with the center around $4$. As our focus is on stationary distributions we give the systems time to achieve stable states. If preferences are not switched on then the network quickly, in less than 100 steps, reaches the stationary state with the exponential degree distribution \cite{Makowiec04}. If the preferences are switched on then the system requires much more time to stabilize.  

The degree distribution is a two-part function in the stationary state. The first part represents properties of vertices which are not preferred, i.e.,  vertices with degree smaller than the threshold ($k<T$). The shape of this part can be approximated by a  power-law with the degree exponent $\gamma$ greater than $3$. The second part of the distribution represents tail properties, i.e., probabilities to meet  vertices with  degrees much greater than the threshold $T$, $k >> T$. The tail decays exponentially. Between these two separated regions there is a  transitional region which can also be characterized  by the power-law dependence on vertex degree. It appears that for the given $T$ there exists  a certain value of $p$ (denoted $p_T$) for which the first region and the transitional region behave according to the common power-law decay with the exponent much smaller than $3$. The distribution at $p_T$ is critical --- for $p > p_T$, the  parts of the distribution become indistinguishable.

On Fig.~\ref{fig2} we present the degree distributions for various values of $p$ and $T=8$. We observe the evolutions  up to 50000 steps and find that after 10000 steps the systems are in  stationary states --- the degree distributions are identical. The distributions are shown on log-log plots for a better visualization of their power-law features. The regions discussed above are easily identified. For the  degrees  smaller than the threshold, here $k<8$, the probability decays in a  polynomial way. For the degrees large --- tail degrees  ($k >50 $), the  probability decay is exponential. The transition region is  between these two parts, i.e., for $ 8<k<50$.  When $p\approx 0.90$  the widest region with the power-law dependence is observed, $2<k<50$. Therefore,  $p_8\approx 0.90 $. Notice, that at $p=0.1$ there are vertices  with the highest degrees  of all  presented  results in Fig.~\ref{fig2}. The network is highly inhomogeneous. However, the highest degrees observed by us at this  threshold value occur  at $p= 0.30$ : $k_{max} \approx 500$. The transition region with this $p$ characterizes by degree exponent $\gamma =0.43 (0.97)$.

\begin{figure}
\includegraphics[width=0.4\textwidth]{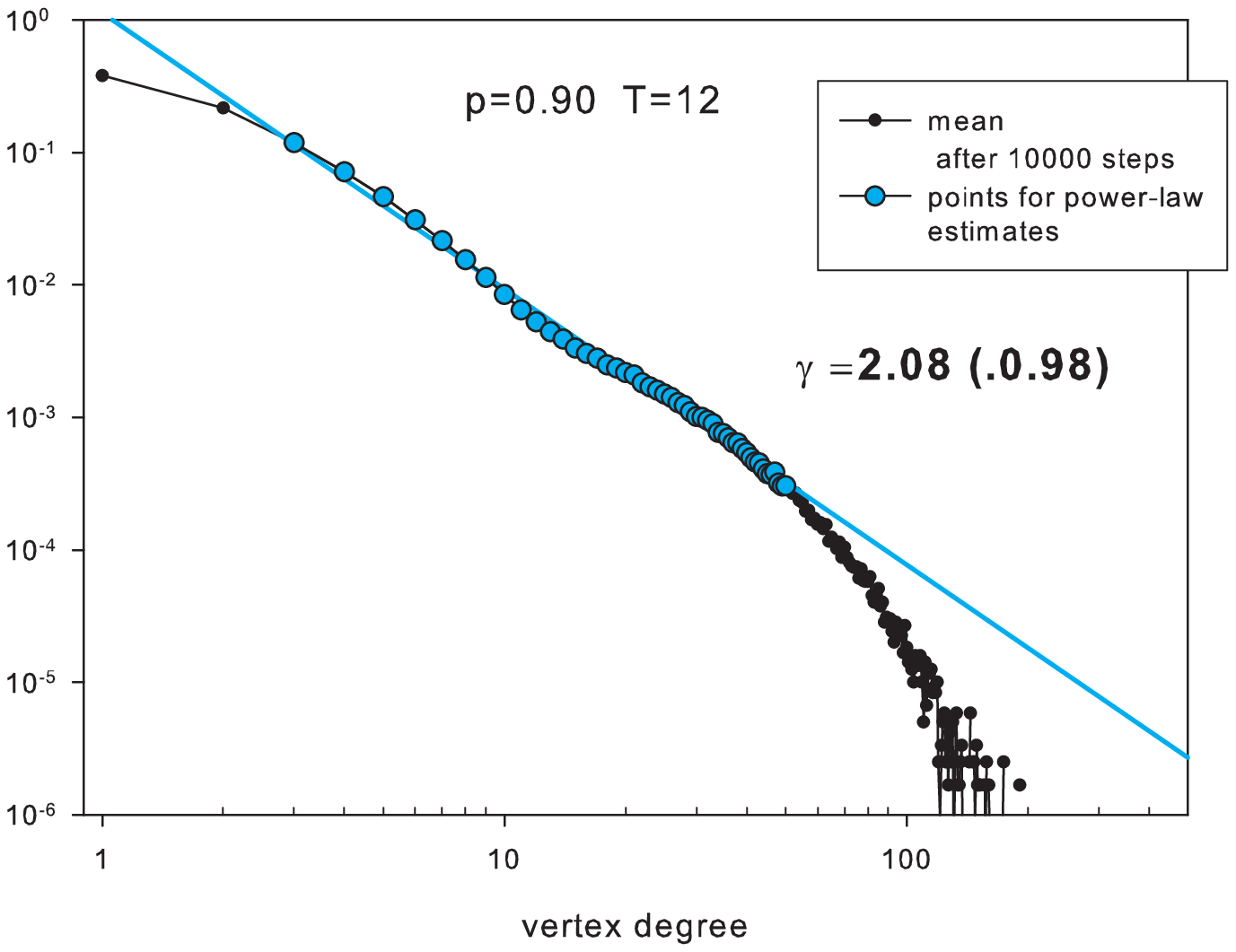}
\includegraphics[width=0.4\textwidth]{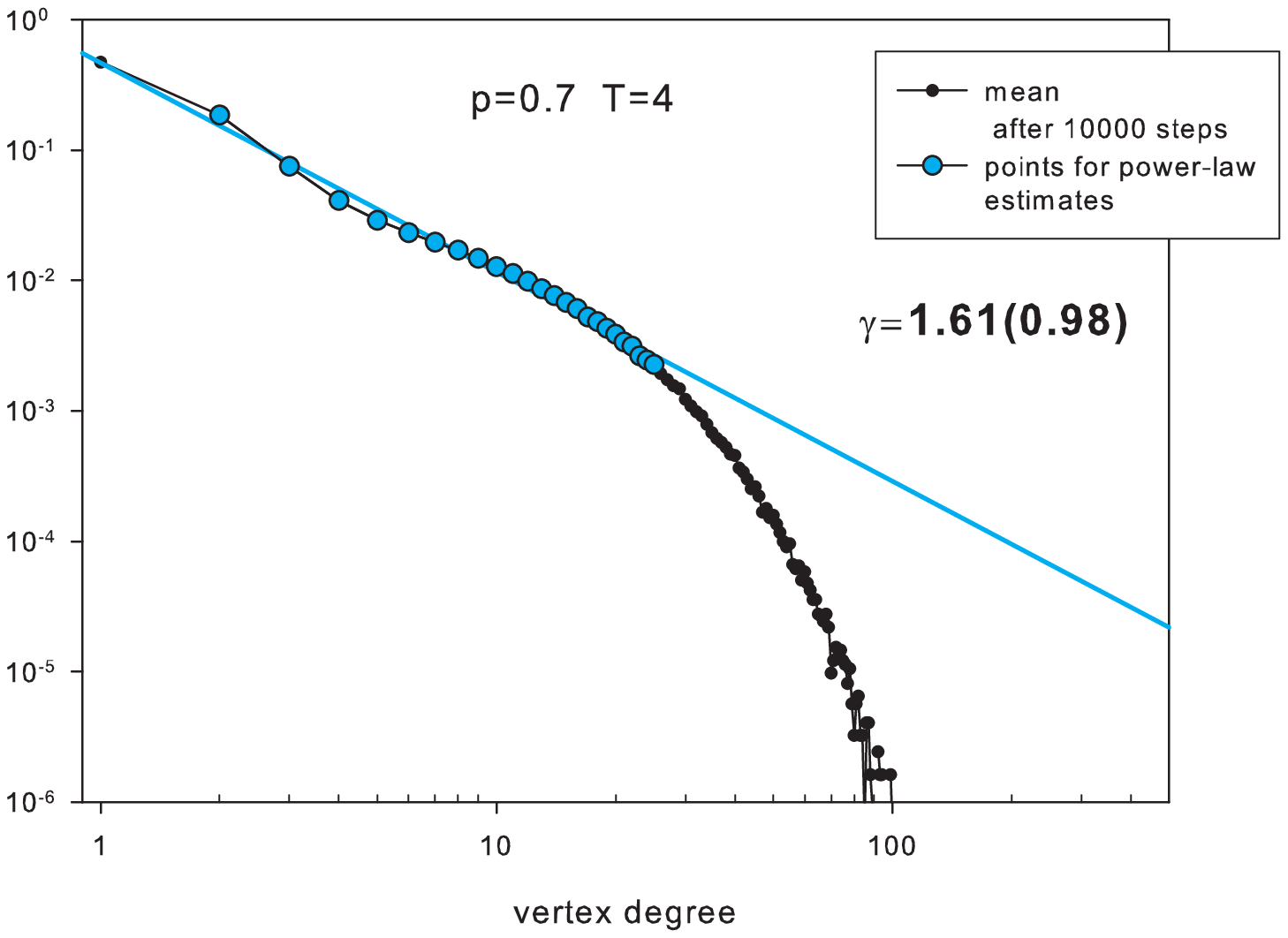}
\caption{\label{fig3} Log-log plots of  critical degree distributions for $T=12$ and $T=4$.}
\end{figure}

\begin{figure}
\includegraphics[width=0.4\textwidth]{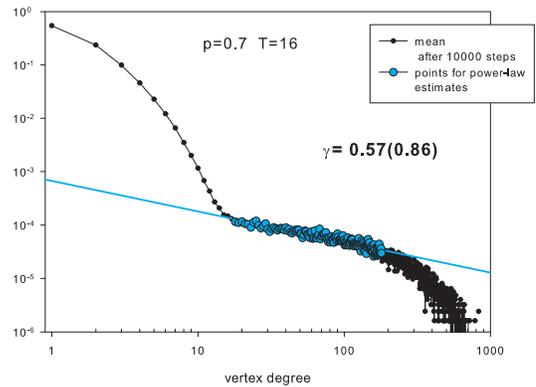}
\caption{\label{fig4} Log-log plot of the widest degree distribution obtained in simulations with  $T=16$.}
\end{figure}

For comparison on  Fig.~\ref{fig3} we show the critical degree distributions for $T=12$ and $T=4$ respectively. We found $p_4\approx 0.7$ and $p_{12}\approx 0.9$. Generally, we observe that  the greater value of $T $, the higher degrees are attainable. In case of $T=12$ and  $p=0.05$  vertices with a degree greater than 600 happens. If $T=16$ then  vertices with $k=1000$ are present on a network, see Fig.~\ref{fig4}. Notice, that in this case the transition region spreads from $k=16$ to almost $k=200$. It implies that with further growth of the threshold  the two-part distribution splits into two separate curves --- there is no transition region.

\section{Model Development}

Watts and Strogatz showed that small worlds captures the best of two graph worlds: the regular and the random. However, there has been no work discussing effect of preferences in rewiring. Our work is a step into this direction. In particular we ask whether small worlds can lead to scale free networks. In the limit of long runs (many time steps)  the evolving network with preferential rewiring may be self-organized into free scale structures. The driving  feature is the requirement to preserve network connected. 

Our work  supports the idea that not growing network may underlie the formation of scale free network. In future attempts to gain more understanding of the organizing principle in complex systems  we will  study other preference functions. In particular,  we will  investigate the modification of the rewiring procedure that restricts the rewiring to a local rule --- the  set of accessible vertices when rewiring a ($from, to$)-edge will be restricted to  the set consisting of nearest-neighbors of a $to$-vertex. By this restriction we will be able to consider new features arising in cellular automata  systems due to the network evolution. Cellular automata are the simplest systems used widely in modeling different natural systems, see \cite{ACRI}.

\begin{acknowledgments}
We wish to acknowledge the support of Polish Ministry of Science and Information Technology Project: PB$\slash$1472$\slash$PO3$\slash$2003$\slash$25
\end{acknowledgments}

%\bibliography{network}% Produces the bibliography via BibTeX.

\end{document}